\documentclass[%
 reprint,
 amsmath,amssymb,
 aps,
 pra,
]{revtex4-2}

\usepackage{graphicx}
\usepackage{dcolumn}
\usepackage{bm}
\usepackage{hyperref}

\usepackage{txfonts}
\usepackage{color}
\usepackage{braket} 
\usepackage{tikz}
\usepackage{quantikz}

\begin{document}

\title{Eigenvalue-invariant transformation of Ising problem for anti-crossing mitigation in quantum annealing}


\author{T. Fujii}
 \email{Toru.Fujii@nikon.com}
\author{K. Komuro}
\author{Y. Okudaira}
\affiliation{%
 Nikon Corporation, 2-15-3, Konan, Minato-ku, 108-6290 Tokyo, Japan 
}
\author{M. Sawada}
\affiliation{
 Nikon Systems Inc., 1-6-3, Nishioi, Shinagawa-ku, 140-0015 Tokyo, Japan
}%

\date{\today}

\begin{abstract}
%
We have proposed the energy landscape transformation of Ising problems (ELTIP), which changes the combination of the state and eigenvalue without changing all the original eigenvalues [arXiv:2202.05927].
We study how the ELTIP affects the anti-crossing between two levels of the ground and first excited states during quantum annealing.
We use a 5-spin maximum-weighted independent set for the problem to numerically investigate the anti-crossing.
For comparison, we introduce a non-stoquastic Hamiltonian that adds antiferromagnetic interaction to the normal transverse magnetic field.
Annealing with the non-stoquastic Hamiltonian is effective for difficult problems.
The non-stoquastic Hamiltonian mitigates the anti-crossing when only the energy gap between the ground state and the first excited state of the final state is small.
When the ELTIP is used, the anti-crossing disappears.
For the problems investigated in this paper, the ELTIP shortens the annealing time to guarantee adiabatic change more than the non-stoquastic Hamiltonian.
\end{abstract}


\maketitle

\section{Introduction}

Quantum annealing \cite{Kadowaki98,Farhi00,Morita08,Albash18,Hauke20} (QA) is a promising method for solving computationally difficult problems using adiabatic changes, and many examples of industrial applications using actual machines have already been reported. Many references are found in Yarnoki et al \cite{Yarkoni22}.
It has also been shown theoretically that the time required for convergence is shorter than simulated annealing \cite{Morita07}.
Quantum logic gate devices are in the ``noisy intermediate-scale quantum'' (NISQ) era and quantum devices of quantum annealer are no exception \cite{Preskill18,Bharti22}.
It is not realistic to perform QA for a time that is theoretically sufficient for convergence.
Instead, various speed-up methods have been proposed to overcome noise limits in both QA and NISQ gate-based devices \cite{Crosson21}.

The difficulty level of the Ising problem is well known.
Various methods using special driving Hamiltonians and/or annealing schedule modification have been proposed to mitigate the difficulty level, and some of them have been demonstrated on the quantum annealing hardware D-Wave  \cite{Dickson12,King17,Koenz19,Marshall19,Chen20,Seki12,Hormozi17,Nishimori17,Ohzeki17}.
Especially, non-stoquastic Hamiltonians \cite{Seki12,Hormozi17,Nishimori17,Ohzeki17} that positively use the quantum property of QA have been shown to be effective for difficult problems \cite{Crosson14}.
Although the non-stoquastic Hamiltonians have a potential to improve the convergence of the QA computing, they are not easy to introduce to QA machines\cite{Mandra17}.

In computer science, it has been shown that the Ising problem with randomly generated coefficients becomes NP-hard \cite{Barahona82}.
However, in the case of limited number of spins which can be simulated by current computers, the probability that it is a difficult problem for non-stoquastic QA was less than 1/1000 in randomly generated 20-spin MAX 2-SAT problems \cite{Crosson14}.
Therefore, designed problems like meticulously parameter-tuned maximum-weighted independent set (MIS) problems \cite{Choi08,Choi20} should be used to analyze mitigation effect in QA.


Theoretically, the required annealing time to obtain the smallest energy state (most optimal answer) can be represented by the adiabatic condition \cite{Morita08}.
The adiabatic condition indicates that the ratio of the maximum time derivation of the Hamiltonian and the minimum of the square of the inverse spectral gap must be small enough for adiabatic computing.
Amin has approximated the adiabatic condition by the energy gap of the ground and first states which mostly affect the adiabatic evolution \cite{Amin09,Amin09b}.
The approximate and rigorous versions of the adiabatic theorems are summarized in the literature \cite{Albash18}.
Choi has given a theorem based on the approximated condition to assess the performance of a QA algorithm \cite{Choi20}, including non-stoquastic Hamiltonians \cite{Choi22}.
The theorem relates the minimum spectral gap (min-gap) and the presence or absence of an anti-crossing during quantum evolution \cite{Farhi00}.
MIS problems \cite{Choi08} are used to justify the theorem.
Here, the anti-crossing is identified as “the local minimum in the plot of the instantaneous energy gap as a function of time” as in the work by Hormozi, et al \cite{Hormozi17}.
Braida and Martiel have expanded Choi's theorem to give a more general expression of anti-crossing \cite{Braida21}.
The study of a specific combinatorial problem called weighted max \textit{k}-clique is shown by using the $\sigma_{y}$ driver.
The above theorems aim to analyze the mechanism of the anti-crossing, not to improve the convergence of QA.

We propose a method to mitigate the approximated adiabatic condition (thus mitigating the anti-crossing) in the quantum evolution of Ising problems with a smaller number of spins.
The method uses ancilla spins to transform the energy landscape of the Ising problems, and is termed the energy landscape transformation of Ising problem (ELTIP) \cite{Fujii22}.
This paper shows that the ELTIP is also explained using controlled-NOT (CNOT), gates.
We numerically show that the ELTIP-transformed problem Hamiltonian is stable for not only the larger degree, but also the small degree of anti-crossing, compared to the non-stoquastic Hamiltonians which are effective only for the larger degree of anti-crossing.
Our method can be realized with the $\sigma_{x}$ driver which is commonly used in QA machines.
Dickson has proposed a method changing the degeneracy of the spectrum of the final Hamiltonian \cite{Dickson11}.
In contrast, the ELTIP does not change degeneracy nor any eigenvalues.
Note that the ELTIP changes the interaction between spins, thus it cannot be used for analysis of physical phenomena of the original quantum evolution.
In this paper, the ELTIP is applied to the 5-spin MIS problems, allowing us to show the effectiveness of the ELTIP more clearly than with fully connected models \cite{Fujii22}.
Moreover, discussions based on the min-gap and the anti-crossing are given.

\section{Annealing time and anti-crossing}
We review the approximate version of the adiabatic theorem and the relation between the min-gap and the anti-crossing \cite{Farhi00,Choi20,Braida21}.

\subsection{Approximate version of adiabatic theorem and Ising problem} 

The annealing time evolution equation is usually 
expressed using the adimensional time $s = t/T$, where $t$ and $T$ denote time and total annealing time, respectively.
A QA problem of an $n$ spin system can be expressed as the following equation \cite{Kadowaki98}:
\begin{eqnarray}
H(s)=(1-s)H_B+sH_P.
\label{eq:Hamiltonian}
\end{eqnarray}
In QA, ${H_B}$ is a transverse magnetic field added by the Pauli $x$-matrix $\sigma_i^x$ as
\begin{eqnarray}
H_B=\sum_{i=0}^{n-1}\sigma_i^x\ 
\label{eq:HB}.
\end{eqnarray}
Equation (\ref{eq:Hamiltonian}) using Eq. (\ref{eq:HB}) is called stoquastic quantum annealing.
Here, the problem Hamiltonian is expressed by the Ising coefficients as
\begin{eqnarray}
H_P = \sum_{i<j} J_{ij}\sigma_i^z\sigma_j^z +  \sum_i h_i\sigma_i^z
\label{eq:HP}.
\end{eqnarray}
Equation (\ref{eq:HP}) consists of the two-spin interaction coefficient ${J_{ij}}$ and the longitudinal magnetic field interaction coefficient ${h_i}$.

It is well known that Quadratic Unconstrained Binary Optimization (QUBO) problems are encoded by Ising Hamiltonians.
The Ising energy can be written by QUBO variables $q_{i} \in \{0, 1\}$ instead of Ising spin variables $\sigma_{i}^{z} \in \{-1, 1\}$:
\begin{eqnarray}
H_{P} &=& 4\sum_{i=0}^{n-1} \sum_{j=0,i<j}^{n-1} J_{ij}q_{i}q_{j} \nonumber \\
&&+ 2\sum_{i=0}^{n-1} \sum_{j=0,i<j}^{n-1} (-J_{ij} + h_{i}) q_{i} + \sum_{i=0}^{n-1} \sum_{j=0,i<j}^{n-1} J_{ij}\nonumber \\
&&- \sum_{i=0}^{n-1} h_{i}
\label{eq:H_QUBO_before}.
\end{eqnarray}
When we use QUBO coefficients $Q_{ij}$ and $b_{i}$ instead of Ising coefficients, $H_{P}$ is rewritten by:
\begin{eqnarray}
H_{P} = Q_{ij}q_{i}q_{j} + b_{i} q_{i} + \text{const.}, 
\label{eq:H_QUBO}
\end{eqnarray}
where $\text{const.}$ is
\begin{eqnarray}
\text{const.} = \sum_{i=0}^{n-1} \sum_{j=0,i<j}^{n-1} J_{ij} - \sum_{i=0}^{n-1} h_{i}
\label{eq:cc}.
\end{eqnarray}

The quantum adiabatic theorem characterizes the sufficiently slow rate required for evolution to move from the initial ground state to the final ground state by giving a lower bound on the total time $T_\text{anneal}$ \cite{Farhi00}:
\begin{eqnarray}
T_{\text{anneal}} \gg \dfrac{\epsilon}{\Delta_{\text{min}}^{2}},
\label{eq:AnnealingTime}
\end{eqnarray}
where $\epsilon = \text{max}_{s} |\braket{E_{1}(s) | \frac{dH}{ds}| E_{0}(s)} |$ and $\Delta_{\text{min}} = \text{min}_{s} \  (E_{1}(s) - E_{0}(s))$ called ``min-gap''.
$E_{0}(s)$ and $E_{1}(s)$ denote the eigenvalues of the ground state $\Ket{E_{0}(s)}$ and the first excited state $\Ket{E_{1}(s)}$, respectively.
They can be directly calculated from $H_{P}$.
On the other hand, $\Ket{\Tilde{E}_{0}(T)}$ is defined as the final ground state obtained by Eq. (\ref{eq:Hamiltonian}), which depends on total annealing time $T$.
When total annealing time $T$ in Eq. (\ref{eq:Hamiltonian}) is longer than $T_{\text{anneal}}$, $\epsilon$ can make $ |\braket{E_{0}(1)| \Tilde{E}_{0}(1)}|$ arbitrarily close to 1 \cite{Farhi00}.
For all of the problems that we simulate in Sec. 4, $\epsilon$ is of order a typical eigenvalue of $H_{P}$ and is not too big, so the size of $T_{\text{anneal}}$ is governed by $\Delta_{\text{min}}^{-2}$.
We use the approximated annealing time $T_{\text{approx}} = \Delta_{\text{min}}^{-2}$ for order-of-magnitude estimate.

\subsection{Anti-crossings} 
Anti-crossings are useful for the analysis of physical phenomena happening during QA, and are the specific behavior of the two lowest eigenvalues when the min-gap occurs around the anti-crossing point $s^{*}$ \cite{Wilkinson89}:
\begin{eqnarray}
E^{\pm}(s) = E(s^{*}) + B(s-s^{*}) \pm \dfrac{1}{2}[ \Delta^{2}_{\text{min}} + A^{2}(s-s^{*})^{2}]^{1/2},
\label{eq:Anti-crossingPoint}
\end{eqnarray}
where $A$ and $B$ are the difference and the mean of the slopes of the asymptotes of the hyperbola, respectively.
An anti-crossing at a first-order phase transition during the evolution occurs in the presence of an exponentially small min-gap.
Choi has suggested a new parametrization of the physical phenomenon to study the occurrence of anti-crossings during adiabatic quantum computation \cite{Choi20}.
Recently, Choi has applied parametrization to non-stoquastic quantum annealing \cite{Choi22}.
Braida and Martiel have suggested another expression of the min-gap that is more general \cite{Braida21}.
In Sec. 4, the min-gaps before/after the ELTIP are discussed.

We express the square of coefficients of the instantaneous eigenstates $\Ket{E_{0}(s)}$ in terms of the eigenstates $\Ket{E_{k} (1)}$ of the final Hamiltonian \cite{Choi20}:
\begin{eqnarray}
a_{k, 0} (s) &=& |\braket{E_{k}(1)|E_{0}(s)}|^{2}. 
\label{eq:ak_bk}
\end{eqnarray}
That is, the square of a coefficient of the instantaneous ground state $a_{0,0}(s) = |\braket{GS|E_{0}(s)}|^{2}$ is the weight (or overlap) of the solution state with the instantaneous ground state at time $s$.
Similarly, $a_{1,0}(s) = |\braket{FS|E_{0}(s)}|^{2}$ is the weight (or overlap) of the first excited state (which possibly corresponds to the
local minima of the problem) with the instantaneous ground state at time $s$.
At $s = 1$, we have $a_{0,0}(1) = 1$ and $a_{1,0}(1) = 0$. 
We use min-gaps to evaluate the effectiveness of the ELTIP.
We also observe the time evolution of $a_{0,0} (s)$ and $a_{1,0} (s)$,  which has the correlation to min-gaps.

\subsection{Non-stoquastic Hamiltonian}
We use the non-stoquastic Hamiltonian \cite{Seki12,Hormozi17,Nishimori17,Ohzeki17} for comparison.
The annealing Hamiltonian including the non-stoquastic Hamiltonian introduced by Seki and Nishimori \cite{Seki12} is expressed as
\begin{eqnarray}
H(s) = s\{\lambda H_{P} + (1 - \lambda) H_{\text{AFF}}\} + (1 - s) H_{B},
\label{eq:nonstq}
\end{eqnarray}
where $H_{\text{AFF}}$ denotes the antiferromagnetic interaction which has a two-spin flip effect:
\begin{eqnarray}
H_{\text{AFF}} = +N \left( \dfrac{1}{N} \sum_{i=0}^{n-1}\sigma_i^x \right)^{2} .
\label{eq:AFF}
\end{eqnarray}
The initial Hamiltonian has $s = 0$ and $\lambda$ is arbitrary, and the final one has $s = \lambda = 1$.

\section{Eigenvalue-invariant transformation of Ising problem } 
\begin{figure}[b]
\centering
\begin{quantikz}[row sep={5mm,between origins}]
\lstick{$k$} 
&\ctrl{1} &\ctrl{2} &\qw      &\ctrl{4} &\qw \\
\lstick[wires=4]{$i\ne k$}
&\targ{1} &\qw      &\qw      &\qw      &\qw \\
&\qw      &\targ{1} &\qw      &\qw      &\qw \\
&         &         &\vdots   &         &\\
&\qw      &\qw      &\qw      &\targ{1} &\qw 
\end{quantikz}
\caption{An eigenvalue-invariant transformation of the Ising problem is depicted in quantum gates.}\label{fig:cnot}
\end{figure}
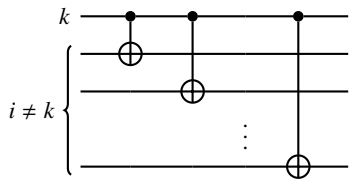

Since eigenvalues of a Hamiltonian are invariant under any unitary transformation, it is possible to choose a basis other than the computational basis to represent solution candidates of the Ising problem, even though it leads to a different physical structure from the original one.
Here, we consider unitary transformations consisting of $n-1$ CNOT gates, shown in Fig. \ref{fig:cnot},
\begin{eqnarray}\label{eq:cnot}
U_k&=&|0_k\rangle \langle 0_k| \otimes \prod_{i\ne k} I_i\nonumber \\
   &+&|1_k\rangle \langle 1_k| \otimes \prod_{i\ne k} \sigma^x_i,~~k=1,\dots,n.
\end{eqnarray}
Due to the anti-commutation of $\sigma^x$ and $\sigma^z$, the Pauli operators are exchanged as follows:
\begin{eqnarray}
U_k \sigma_k^z \sigma_i^z U_k^\dagger &=& \sigma_i^z~~~\text{and}\\
U_k \sigma_i^z U_k^\dagger &=& \sigma_k^z \sigma_i^z~~~\text{for}~i\ne k,
\end{eqnarray}
and leaves the following unchanged:
\begin{eqnarray}
U_k \sigma_k^z U_k^\dagger &=& \sigma_k^z,\\
U_k \sigma_i^z \sigma_j^z U_k^\dagger &=& \sigma_i^z \sigma_j^z~~~\text{for}~i,j\ne k.
\end{eqnarray}
Applying the unitary transformations Eq. (\ref{eq:cnot}) to the original Ising Hamiltonian Eq. (\ref{eq:HP}) yields $n$ Ising Hamiltonians $H_k=U_k H U_k^\dagger,~ k=1,\dots,n$, i.e.
\begin{eqnarray}\label{eq:eltip}
H_k&=&\sum_{i,j\ne k} J_{ij} \sigma_i^z\sigma_j^z +\sum_{i\ne k}J_{ik}\sigma_i^z
+\sum_{i\ne k}h_i \sigma_i^z \sigma_k^z +h_k \sigma_k^z,
\end{eqnarray}
where the role of coefficients of second order terms $\{J_{ik}\}$ and first order terms $\{h_i\}$ for $i\ne k$ are exchanged \cite{Fujii22}.
By this transformation, the energy gap of the ground state and the first excited state, during the adiabatic evolution or quantum annealing process, can be altered. Therein lies the potential to alleviate the problem.
Due to the unitary invariance, once one of these problems represented by $H_k$ is solved to give the ground state $|GS_k\rangle$, we have the ground state of the original problem by (classical) calculation of $|GS\rangle= U_k^\dagger|GS_k\rangle$.

Note that repetition of the CNOT Eq. (\ref{eq:cnot}) yields $U_j U_i U_j = \text{SWAP}_{ij}$, which means the repeated application of the CNOT  results in at most a single $U_{k}$ and worthless SWAPs.

\section{Time evolution of problems with anti-crossing}

\subsection{Design of problems with anti-crossing by maximum-weighted independent set (MIS)}
The connection between local minima in the problem Hamiltonian designed by MIS and first order quantum phase transitions (QPT) during an adiabatic quantum computation was investigated by Amin and Choi \cite{Amin09,Amin09b}. QPT can be easily generated by using the balance of the MIS to make the eigenvalues of the ground state and the first excited state closer together and to separate the eigenvalues of the other states \cite{Amin09,Amin09b}.
Therefore, we use the MIS to design problems that cause QPT and use it for anti-crossing mitigation effect evaluation.

\begin{figure*}[t]
  \centering
  \includegraphics[width=140mm]{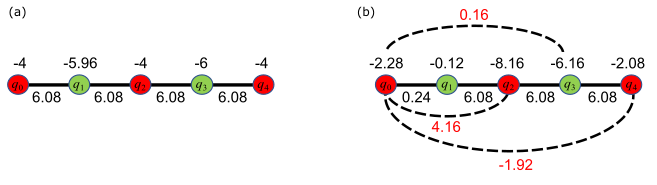}
  \caption{Problems to be simulated in Sec. 4. (a) the original MIS problem and (b) the transformed problem by ELTIP of $H_{0}$ in Eq. (\ref{eq:eltip}). The problems are converted to the QUBO form. The values above the circles denote the weights $b_{i}$ and the other values near the edges denote $Q_{ij}$.}
  \label{fig:MIS}
\end{figure*}

We use the 5-spin MIS problems \cite{Choi20} to compare the effectiveness of the non-stoquastic Hamiltonian and the ELTIP because of the design simplicity and good visibility of numerical results.
An MIS problem shown in Fig. \ref{fig:MIS}(a) is used to demonstrate the effectiveness of the ELTIP.
\begin{table*}[t]
\centering
\caption{
QUBO and Ising coefficients.
}
\label{tab:table1}
\scalebox{1.0}{
\begin{tabular}{l|rrrrrrrrrrrrrrr}
\hline
Coeff. type &
$Q_{01}/J_{01}$&
$Q_{02}/J_{02}$&
$Q_{03}/J_{03}$&
$Q_{04}/J_{04}$&
$Q_{12}/J_{12}$&
$Q_{13}/J_{13}$&
$Q_{14}/J_{14}$&
$Q_{23}/J_{23}$&
$Q_{24}/J_{24}$&
$Q_{34}/J_{34}$&
$b_{0}/h_{0}$&
$b_{1}/h_{1}$&
$b_{2}/h_{2}$&
$b_{3}/h_{3}$&
$b_{4}/h_{4}$
\\
\hline \hline
QUBO & 6.08 & 0 & 0 & 0 & 6.08 & 0 & 0 & 6.08 & 0 & 6.08 & -4 & -5.96 & -4 & -6 & -4
\\
Ising & 1.52 & 0 & 0 & 0 & 1.52& 0 & 0 & 1.52 & 0 & 1.52 & -0.48 & 0.06 & 1.04 & 0.04 & -0.48
\\
Ising $H_{0}$ & 0.06 & 1.04 & 0.04 & -0.48 & 1.52 & 0 & 0 & 1.52 & 0 & 1.52 & -0.48 & 1.52 & 0 & 0 & 0
\\
QUBO $H_{0}$ & 0.24 & 4.16 & 0.16 & -1.92 & 6.08 & 0 & 0 & 6.08 & 0 & 6.08 & -2.28 & -0.12 & -8.16 & -6.16 & -2.08
\\
\hline
\end{tabular}
}
\end{table*}

\begin{figure*}[t]
  \centering
  \includegraphics[width=100mm]{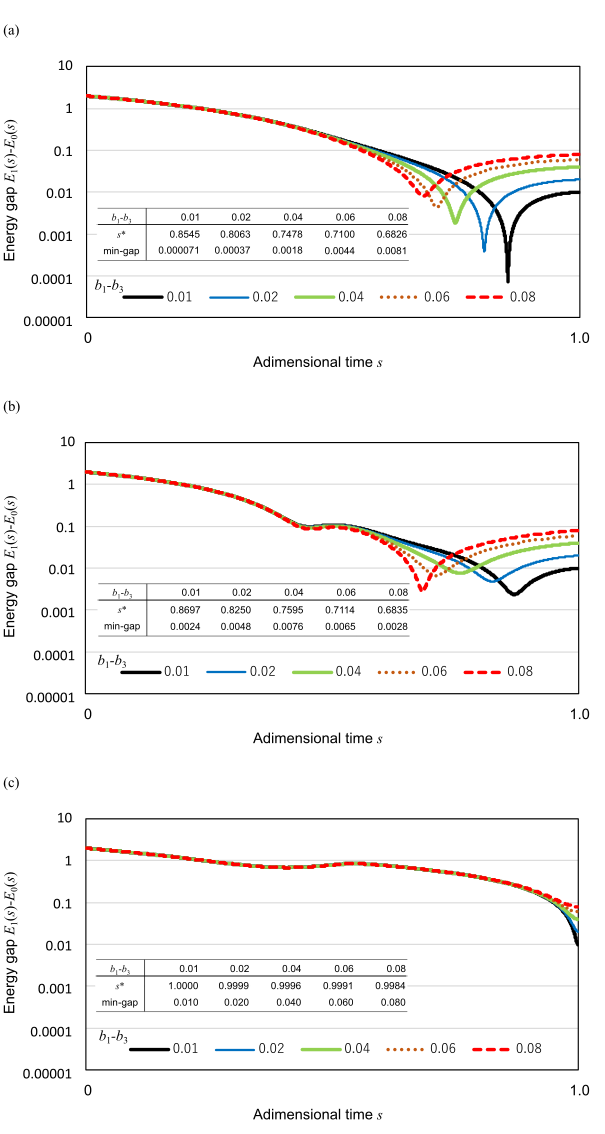}
  \caption{Numerical results of the energy gap. (a) the original MIS problem, (b) the problem by the non-stoquastic Hamiltonian, and (c) the transformed problem by the ELTIP.}
  \label{fig:NumericalResults}
\end{figure*}
\begin{figure*}[t]
  \centering
  \includegraphics[width=140mm]{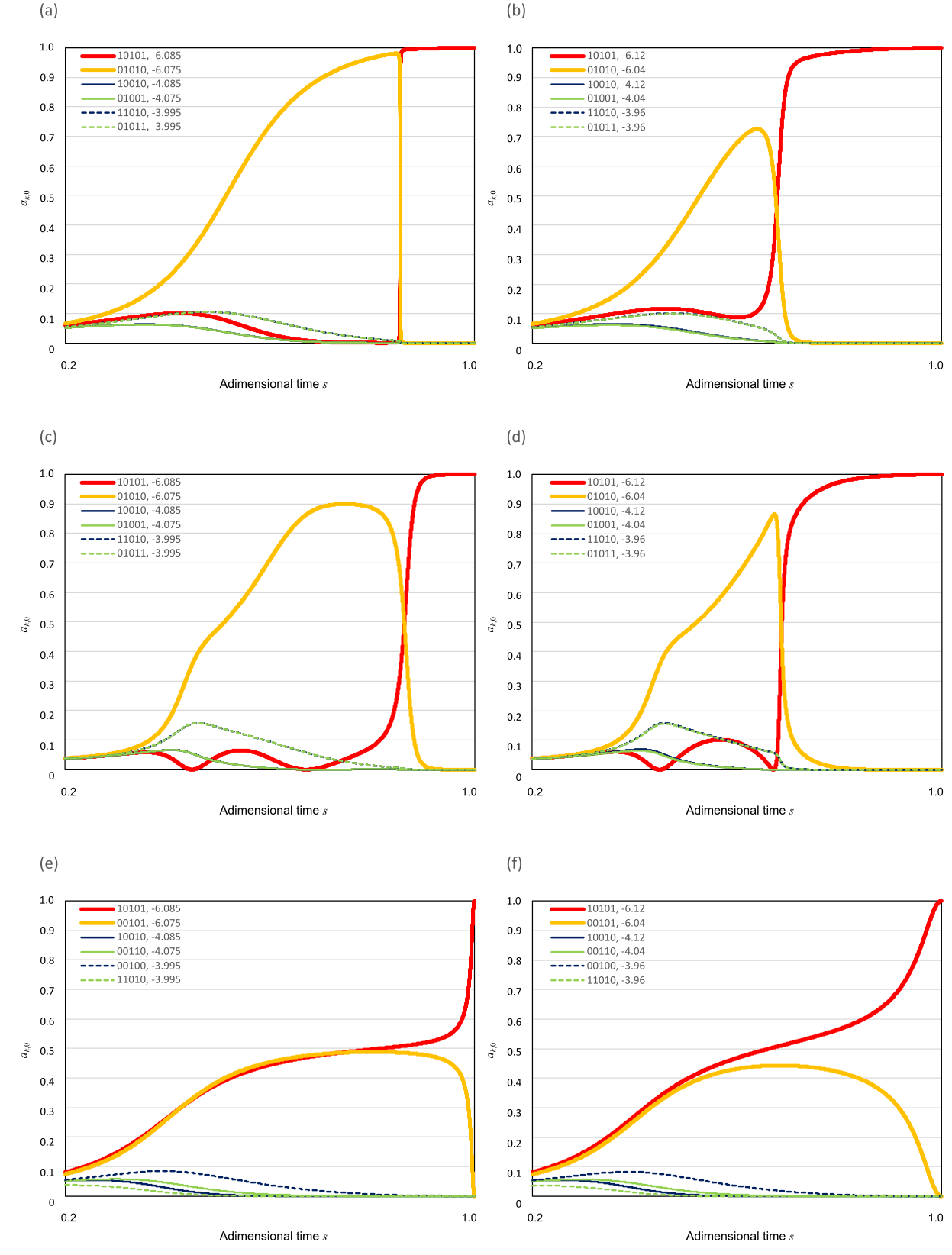}
  \caption{Numerical results of the instantaneous ground state represented by the final eigenstates with adinmensional time $s = [0.2, 1]$ in a horizontal axis. A vertical axis denotes square of coefficients of the final eigenstates for $k=0, 1, ..., 5$ as is shown in Ref. \cite{Choi20}. Left and right are the instantaneous ground states with $E_{1}(1)-E_{3}(1) = \Delta b_{1,3} = 0.01\ \text{and}\ 0.08$, respectively. (a) (b) the instantaneous ground state during the stoquastic quantum annealing in the original MIS problem, (c) (d) the instantaneous ground state during the non-stoquastic quantum annealing, and (e) (f) the instantaneous ground state during the stoquastic quantum annealing in the Ising Hamiltonian transformed by the ELTIP.}
  \label{fig:NumericalResultsEnergy}
\end{figure*}
%
%
As an MIS problem, only the weights $b_{i}$ are important, but the coupling constants $Q_{ij}$ of 6.08 in Fig. \ref{fig:MIS}(a) are not important.
In other words, the problem with the other coupling constants can be regarded as the same MIS.
However, whole energy landscapes are changed by the coupling constants, which affects QA time evolution.
Therefore, the coupling constants must be fixed for the evaluation of QA time evolution and conversion.
Figure \ref{fig:MIS}(a) is designed to generate the anti-crossing between the ground and the first states, referred to in Ref \cite{Choi20}.
Note that Ref \cite{Choi20} describes the MIS problems by the QUBO form, therefore Fig. \ref{fig:MIS}(a) is denoted by the QUBO form.
The ELTIP of $H_{0}=U_0 H U_0^\dagger$ in Eq. (\ref{eq:eltip}) is applied to the problem shown in Fig. \ref{fig:MIS}(a), and a transformed problem is shown in Fig. \ref{fig:MIS}(b).
Compared with the original problem in Fig. \ref{fig:MIS}(a), the transformed problem in Fig. \ref{fig:MIS}(b) has additional edges.
From the viewpoint of obtaining the lowest eigenvalue and state, the problem transformed by the ELTIP is equivalent to the problem before transformation.
Note that the ELTIP changes the physical picture.
Therefore, the ELTIP can reduce the difficulty of solving the Ising problems with QA, while it cannot be used to analyze the physical phenomena in the original problems.

The QUBO and corresponding Ising coefficients with the ELTIP process are summarized in Table. \ref{tab:table1}.
The ELTIP can be used for Ising problems.
Therefore, once the original QUBO problem is written in the Ising problem (the first line to the second line in Table. \ref{tab:table1}), then the ELTIP is applied to the Ising problem (the second line to the third line in Table. \ref{tab:table1}).
Finally, the Ising coefficient with the ELTIP is converted to the QUBO form (the third line to the fourth line in Table. \ref{tab:table1}).

Figure \ref{fig:MIS}(a) and the first line in Table. \ref{tab:table1}, are an example in the case of $\Delta b_{1, 3} = b_{1} - b_{3} = -5.96 - (-6) = 0.04$.
It is well known that the energy landscape affects the convergence of annealing.
By changing $\Delta b_{1, 3}$, the energy landscape almost remains, but the energy gap between the final ground state and the final first excited state is tunable.
Problems of $\Delta b_{1, 3} = 0.01, 0.02, 0.04, 0.06, 0.08$ will be simulated in the next subsection.

\subsection{Min-gaps of MIS problems}

The energy gaps of the problems are shown in Fig. \ref{fig:NumericalResults}, where the energy gaps are $E_{1}(s)-E_{0}(s)$.
We compare time evolution of the stoquastic quantum annealing, the non-stoquastic quantum annealing, and the stoquastic quantum annealing with the ELTIP transformation.
$s^{*}$ is defined as $s$ when the min-gap occurs. 
$E_{1}(1)-E_{0}(1)$ is called the final gap, which equals to $\Delta b_{1,3}$.
In the case of the stoquastic quantum annealing, the reduction of min-gap is much larger than that of the final gap.
As the final gap decreases, $s^{*}$ approaches the final state.
In the case of the non-stoquastic quantum annealing, the reduction of min-gap is not proportional to that of final gap.
The non-stoquastic quantum annealing effectively mitigates the anti-crossing for the smallest final gap.
However, as the final gap increases, the mitigation of the anti-crossing is reduced.
For the largest final gap in our case, the anti-crossing is enhanced, although $s^{*}$ is not so changed.
As shown in Fig. \ref{fig:NumericalResults}(c), the ELTIP eliminates the anti-crossing for all the final gaps.
The min-gap is achieved near the end of the annealing, in all the final gaps.

\subsection{Time evolution of instantaneous ground state}
According to Eq. (\ref{eq:AnnealingTime}), the approximated annealing time $T_{\text{approx}} = \Delta_{\text{min}}^{-2}$ determines the lower limit of the annealing time.
We compared the time evolution of $a_{k,0}(s)\  (k=0, 1, ..., 5)$ in the original and the transformed problems as shown in Fig. \ref{fig:NumericalResultsEnergy}.
Left and right plots are the instantaneous ground state represented by the final eigenstates with the final gap of 0.01 and 0.08, respectively.
Figures \ref{fig:NumericalResultsEnergy}(a) and \ref{fig:NumericalResultsEnergy}(b) are the stoquastic annealing time evolution.
For both energy gaps, near $s^{*}$ where anti-crossing occurs, the population ratios of the ground state and the first excited state are rapidly switched.
The switching speed, namely spin-polarity flipping speed of the final gap 0.01 is much faster than that of 0.08.
A wider gap weakens the flipping speed which is one of the characteristics of QPT.
In the case of the non-stoquastic Hamiltonian in Fig. \ref{fig:NumericalResultsEnergy}(c), the rapid transition from the first excited state to the ground state in Fig. \ref{fig:NumericalResultsEnergy}(a) is obviously mitigated.
On the other hand, in Fig. \ref{fig:NumericalResultsEnergy}(d), the slope around $s^{*}$ becomes steeper than that of Fig. \ref{fig:NumericalResultsEnergy}(b) of the stoquastic annealing.
$T_{\text{approx}}$ defined in Sec. 2.1 in the case of the non-stoquastic Hamiltonian with $\Delta b_{1,3} = 0.08$ is 10 times longer than the stoquastic Hamiltonian, and is almost comparable to the case when $\Delta b_{1,3} = 0.01$ of the non-stoquastic Hamiltonian.
The ELTIP shortens $T_{\text{approx}}$ of all the problems where the anti-crossing occurs.
As a result, the ELTIP eliminates the rapid population inversion between the ground state and the first excited state which is observed in QPT.
The ELTIP reduced the approximated annealing time $T_{\text{approx}}$ by a factor of $10^{4}$ with $\Delta b_{1,3} = 0.01$ and $10^{2}$ with $\Delta b_{1,3} = 0.08$.
The rest of four ELTIPs $H_{1}$ to $H_{4}$ shows similar results.

\section{Summary}
We have proposed a method called ELTIP to mitigate the anti-crossing in QA, and have investigated its effectiveness numerically.
The 5-spin QUBO problem using the MIS allows the adjustment of the size of the gap between the ground state and the first excited state while keeping the entire energy landscape.
We also introduce a non-stoquastic Hamiltonian to the normal transverse magnetic field.
Compared to the stoquastic Hamiltonian, the non-stoquastic Hamiltonian shortens the approximated annealing time $T_{\text{approx}}$ by a factor of $10^3$ with the MIS problem of the final gap $E_{1}(1)-E_{0}(1) = 0.01$ which equals to $\Delta b_{1,3} = 0.01$.
As the final gap of the original problem becomes larger, that is, as the problem becomes easier, the effect of shortening the annealing time is reduced.
It can be found that the non-stoquastic Hamiltonian narrows the min-gap more than 10 times in the problems with the largest final gap of 0.08.
The ELTIP eliminates the anti-crossing and also reduces the transition rate from the instant first excited state to the instant ground state.
The ELTIP shortens the approximated annealing time $T_{\text{approx}}$ by a factor of from $10^{2}$ to $10^{4}$ in all the cases.
Due to the unitary invariance, once Ising problems transformed by the ELTIP $H_k$ is solved to give the ground state $|GS_k\rangle$, we have the ground state of the original problem by calculation of $|GS\rangle= U_k^\dagger|GS_k\rangle$.
The ELTIP is effective for practical optimization problems embedded in QUBO or Ising problems, because only the lowest state and eigenvalues of the problems are important.
In this paper, a clear effect was observed in the 5-spin MIS problem, where the min-gap constriction effect is easy to see.
In the future, a variety of trials for problems with different energy landscapes and large spin counts will be required to confirm the effectiveness of our proposed method.

\section*{Acknowledgment} 
The authors thank Haruki Maeda for usefull discussions on anti-crossings in QA problems.
The authors thank Juan Ivaldi for his helpful advice on understanding and representing physical phenomena.

\bibliography{references}

\begin{thebibliography}{31}%
\makeatletter
\providecommand \@ifxundefined [1]{%
 \@ifx{#1\undefined}
}%
\providecommand \@ifnum [1]{%
 \ifnum #1\expandafter \@firstoftwo
 \else \expandafter \@secondoftwo
 \fi
}%
\providecommand \@ifx [1]{%
 \ifx #1\expandafter \@firstoftwo
 \else \expandafter \@secondoftwo
 \fi
}%
\providecommand \natexlab [1]{#1}%
\providecommand \enquote  [1]{``#1''}%
\providecommand \bibnamefont  [1]{#1}%
\providecommand \bibfnamefont [1]{#1}%
\providecommand \citenamefont [1]{#1}%
\providecommand \href@noop [0]{\@secondoftwo}%
\providecommand \href [0]{\begingroup \@sanitize@url \@href}%
\providecommand \@href[1]{\@@startlink{#1}\@@href}%
\providecommand \@@href[1]{\endgroup#1\@@endlink}%
\providecommand \@sanitize@url [0]{\catcode `\\12\catcode `\$12\catcode
  `\&12\catcode `\#12\catcode `\^12\catcode `\_12\catcode `\%12\relax}%
\providecommand \@@startlink[1]{}%
\providecommand \@@endlink[0]{}%
\providecommand \url  [0]{\begingroup\@sanitize@url \@url }%
\providecommand \@url [1]{\endgroup\@href {#1}{\urlprefix }}%
\providecommand \urlprefix  [0]{URL }%
\providecommand \Eprint [0]{\href }%
\providecommand \doibase [0]{https://doi.org/}%
\providecommand \selectlanguage [0]{\@gobble}%
\providecommand \bibinfo  [0]{\@secondoftwo}%
\providecommand \bibfield  [0]{\@secondoftwo}%
\providecommand \translation [1]{[#1]}%
\providecommand \BibitemOpen [0]{}%
\providecommand \bibitemStop [0]{}%
\providecommand \bibitemNoStop [0]{.\EOS\space}%
\providecommand \EOS [0]{\spacefactor3000\relax}%
\providecommand \BibitemShut  [1]{\csname bibitem#1\endcsname}%
\let\auto@bib@innerbib\@empty
\bibitem [{\citenamefont {Kadowaki}\ and\ \citenamefont
  {Nishimori}(1998)}]{Kadowaki98}%
  \BibitemOpen
  \bibfield  {author} {\bibinfo {author} {\bibfnamefont {T.}~\bibnamefont
  {Kadowaki}}\ and\ \bibinfo {author} {\bibfnamefont {H.}~\bibnamefont
  {Nishimori}},\ }\href@noop {} {\bibfield  {journal} {\bibinfo  {journal}
  {Phys.\ Rev.\ E}\ }\textbf {\bibinfo {volume} {58}},\ \bibinfo {pages} {5355}
  (\bibinfo {year} {1998})}\BibitemShut {NoStop}%
\bibitem [{\citenamefont {Farhi}\ \emph {et~al.}(2000)\citenamefont {Farhi},
  \citenamefont {Goldstone}, \citenamefont {Gutmann},\ and\ \citenamefont
  {Sipser}}]{Farhi00}%
  \BibitemOpen
  \bibfield  {author} {\bibinfo {author} {\bibfnamefont {E.}~\bibnamefont
  {Farhi}}, \bibinfo {author} {\bibfnamefont {J.}~\bibnamefont {Goldstone}},
  \bibinfo {author} {\bibfnamefont {S.}~\bibnamefont {Gutmann}},\ and\ \bibinfo
  {author} {\bibfnamefont {M.}~\bibnamefont {Sipser}},\ }\href@noop {}
  {}\bibinfo {howpublished} {e-print arXiv:quant-ph/0001106} (\bibinfo {year}
  {2000})\BibitemShut {NoStop}%
\bibitem [{\citenamefont {Morita}\ and\ \citenamefont
  {Nishimori}(2008)}]{Morita08}%
  \BibitemOpen
  \bibfield  {author} {\bibinfo {author} {\bibfnamefont {S.}~\bibnamefont
  {Morita}}\ and\ \bibinfo {author} {\bibfnamefont {H.}~\bibnamefont
  {Nishimori}},\ }\href@noop {} {\bibfield  {journal} {\bibinfo  {journal} {J.
  Math. Phys.}\ }\textbf {\bibinfo {volume} {49}},\ \bibinfo {pages} {125210}
  (\bibinfo {year} {2008})}\BibitemShut {NoStop}%
\bibitem [{\citenamefont {Albash}\ and\ \citenamefont
  {Lidar}(2018)}]{Albash18}%
  \BibitemOpen
  \bibfield  {author} {\bibinfo {author} {\bibfnamefont {T.}~\bibnamefont
  {Albash}}\ and\ \bibinfo {author} {\bibfnamefont {D.~A.}\ \bibnamefont
  {Lidar}},\ }\href@noop {} {\bibfield  {journal} {\bibinfo  {journal} {Rev.
  Mod. Phys.}\ }\textbf {\bibinfo {volume} {90}},\ \bibinfo {pages} {015002}
  (\bibinfo {year} {2018})}\BibitemShut {NoStop}%
\bibitem [{\citenamefont {Hauke}\ \emph {et~al.}(2020)\citenamefont {Hauke},
  \citenamefont {Katzgraber}, \citenamefont {Lechner}, \citenamefont
  {Nishimori},\ and\ \citenamefont {Oliver}}]{Hauke20}%
  \BibitemOpen
  \bibfield  {author} {\bibinfo {author} {\bibfnamefont {P.}~\bibnamefont
  {Hauke}}, \bibinfo {author} {\bibfnamefont {H.~G.}\ \bibnamefont
  {Katzgraber}}, \bibinfo {author} {\bibfnamefont {W.}~\bibnamefont {Lechner}},
  \bibinfo {author} {\bibfnamefont {H.}~\bibnamefont {Nishimori}},\ and\
  \bibinfo {author} {\bibfnamefont {W.~D.}\ \bibnamefont {Oliver}},\
  }\href@noop {} {\bibfield  {journal} {\bibinfo  {journal} {Rep. Prog. Phys.}\
  }\textbf {\bibinfo {volume} {83}},\ \bibinfo {pages} {054401} (\bibinfo
  {year} {2020})}\BibitemShut {NoStop}%
\bibitem [{\citenamefont {Yarkoni}\ \emph {et~al.}(2022)\citenamefont
  {Yarkoni}, \citenamefont {Raponi}, \citenamefont {Back},\ and\ \citenamefont
  {Schmitt}}]{Yarkoni22}%
  \BibitemOpen
  \bibfield  {author} {\bibinfo {author} {\bibfnamefont {S.}~\bibnamefont
  {Yarkoni}}, \bibinfo {author} {\bibfnamefont {E.}~\bibnamefont {Raponi}},
  \bibinfo {author} {\bibfnamefont {T.}~\bibnamefont {Back}},\ and\ \bibinfo
  {author} {\bibfnamefont {S.}~\bibnamefont {Schmitt}},\ }\href@noop {}
  {}\bibinfo {howpublished} {e-print arXiv:2112.07491} (\bibinfo {year}
  {2022})\BibitemShut {NoStop}%
\bibitem [{\citenamefont {Morita}\ and\ \citenamefont
  {Nishimori}(2007)}]{Morita07}%
  \BibitemOpen
  \bibfield  {author} {\bibinfo {author} {\bibfnamefont {S.}~\bibnamefont
  {Morita}}\ and\ \bibinfo {author} {\bibfnamefont {H.}~\bibnamefont
  {Nishimori}},\ }\href@noop {} {\bibfield  {journal} {\bibinfo  {journal} {J.
  Phys. Soc. Jpn.}\ }\textbf {\bibinfo {volume} {76}},\ \bibinfo {pages}
  {064002} (\bibinfo {year} {2007})}\BibitemShut {NoStop}%
\bibitem [{\citenamefont {Preskill}(2018)}]{Preskill18}%
  \BibitemOpen
  \bibfield  {author} {\bibinfo {author} {\bibfnamefont {J.}~\bibnamefont
  {Preskill}},\ }\href@noop {} {\bibfield  {journal} {\bibinfo  {journal}
  {Quantum}\ }\textbf {\bibinfo {volume} {2}},\ \bibinfo {pages} {79} (\bibinfo
  {year} {2018})}\BibitemShut {NoStop}%
\bibitem [{\citenamefont {Bharti}\ \emph {et~al.}(2022)\citenamefont {Bharti},
  \citenamefont {Cervera-Lierta}, \citenamefont {Kyaw}, \citenamefont {Haug},
  \citenamefont {Alperin-Lea}, \citenamefont {Anand}, \citenamefont {Degroote},
  \citenamefont {Heimonen}, \citenamefont {Kottmann}, \citenamefont {Menke},
  \citenamefont {Mok}, \citenamefont {Sim}, \citenamefont {Kwek},\ and\
  \citenamefont {Aspuru-Guzik}}]{Bharti22}%
  \BibitemOpen
  \bibfield  {author} {\bibinfo {author} {\bibfnamefont {K.}~\bibnamefont
  {Bharti}}, \bibinfo {author} {\bibfnamefont {A.}~\bibnamefont
  {Cervera-Lierta}}, \bibinfo {author} {\bibfnamefont {T.~H.}\ \bibnamefont
  {Kyaw}}, \bibinfo {author} {\bibfnamefont {T.}~\bibnamefont {Haug}}, \bibinfo
  {author} {\bibfnamefont {S.}~\bibnamefont {Alperin-Lea}}, \bibinfo {author}
  {\bibfnamefont {A.}~\bibnamefont {Anand}}, \bibinfo {author} {\bibfnamefont
  {M.}~\bibnamefont {Degroote}}, \bibinfo {author} {\bibfnamefont
  {H.}~\bibnamefont {Heimonen}}, \bibinfo {author} {\bibfnamefont {J.~S.}\
  \bibnamefont {Kottmann}}, \bibinfo {author} {\bibfnamefont {T.}~\bibnamefont
  {Menke}}, \bibinfo {author} {\bibfnamefont {W.-K.}\ \bibnamefont {Mok}},
  \bibinfo {author} {\bibfnamefont {S.}~\bibnamefont {Sim}}, \bibinfo {author}
  {\bibfnamefont {L.-C.}\ \bibnamefont {Kwek}},\ and\ \bibinfo {author}
  {\bibfnamefont {A.}~\bibnamefont {Aspuru-Guzik}},\ }\href@noop {} {\bibfield
  {journal} {\bibinfo  {journal} {Rev. Mod. Phys.}\ }\textbf {\bibinfo {volume}
  {94}},\ \bibinfo {pages} {79} (\bibinfo {year} {2022})}\BibitemShut {NoStop}%
\bibitem [{\citenamefont {Crosson}\ and\ \citenamefont
  {Lider}(2021)}]{Crosson21}%
  \BibitemOpen
  \bibfield  {author} {\bibinfo {author} {\bibfnamefont {E.~J.}\ \bibnamefont
  {Crosson}}\ and\ \bibinfo {author} {\bibfnamefont {D.~A.}\ \bibnamefont
  {Lider}},\ }\href@noop {} {\bibfield  {journal} {\bibinfo  {journal} {Nat.
  Rev. Phys.}\ }\textbf {\bibinfo {volume} {3}},\ \bibinfo {pages} {466}
  (\bibinfo {year} {2021})}\BibitemShut {NoStop}%
\bibitem [{\citenamefont {Dickson}\ and\ \citenamefont
  {Amin}(2012)}]{Dickson12}%
  \BibitemOpen
  \bibfield  {author} {\bibinfo {author} {\bibfnamefont {N.~G.}\ \bibnamefont
  {Dickson}}\ and\ \bibinfo {author} {\bibfnamefont {M.~H.}\ \bibnamefont
  {Amin}},\ }\href@noop {} {\bibfield  {journal} {\bibinfo  {journal} {Phys.
  Rev. A}\ }\textbf {\bibinfo {volume} {85}},\ \bibinfo {pages} {032303}
  (\bibinfo {year} {2012})}\BibitemShut {NoStop}%
\bibitem [{\citenamefont {Lanting}\ \emph {et~al.}(2017)\citenamefont
  {Lanting}, \citenamefont {King}, \citenamefont {Evert},\ and\ \citenamefont
  {Hoskinson}}]{King17}%
  \BibitemOpen
  \bibfield  {author} {\bibinfo {author} {\bibfnamefont {T.}~\bibnamefont
  {Lanting}}, \bibinfo {author} {\bibfnamefont {A.~D.}\ \bibnamefont {King}},
  \bibinfo {author} {\bibfnamefont {B.}~\bibnamefont {Evert}},\ and\ \bibinfo
  {author} {\bibfnamefont {E.}~\bibnamefont {Hoskinson}},\ }\href@noop {}
  {\bibfield  {journal} {\bibinfo  {journal} {Phys. Rev. A}\ }\textbf {\bibinfo
  {volume} {96}},\ \bibinfo {pages} {042322} (\bibinfo {year}
  {2017})}\BibitemShut {NoStop}%
\bibitem [{\citenamefont {Könz}\ \emph {et~al.}(2019)\citenamefont {Könz},
  \citenamefont {Mazzola}, \citenamefont {Ochoa}, \citenamefont {Katzgraber},\
  and\ \citenamefont {Troyer}}]{Koenz19}%
  \BibitemOpen
  \bibfield  {author} {\bibinfo {author} {\bibfnamefont {M.~S.}\ \bibnamefont
  {Könz}}, \bibinfo {author} {\bibfnamefont {G.}~\bibnamefont {Mazzola}},
  \bibinfo {author} {\bibfnamefont {A.~J.}\ \bibnamefont {Ochoa}}, \bibinfo
  {author} {\bibfnamefont {H.~G.}\ \bibnamefont {Katzgraber}},\ and\ \bibinfo
  {author} {\bibfnamefont {M.}~\bibnamefont {Troyer}},\ }\href@noop {}
  {\bibfield  {journal} {\bibinfo  {journal} {Phys. Rev. A}\ }\textbf {\bibinfo
  {volume} {100}},\ \bibinfo {pages} {030303(R)} (\bibinfo {year}
  {2019})}\BibitemShut {NoStop}%
\bibitem [{\citenamefont {Marshall}\ \emph {et~al.}(2019)\citenamefont
  {Marshall}, \citenamefont {Venturelli}, \citenamefont {Hen},\ and\
  \citenamefont {Rieffel}}]{Marshall19}%
  \BibitemOpen
  \bibfield  {author} {\bibinfo {author} {\bibfnamefont {J.}~\bibnamefont
  {Marshall}}, \bibinfo {author} {\bibfnamefont {D.}~\bibnamefont
  {Venturelli}}, \bibinfo {author} {\bibfnamefont {I.}~\bibnamefont {Hen}},\
  and\ \bibinfo {author} {\bibfnamefont {E.~G.}\ \bibnamefont {Rieffel}},\
  }\href@noop {} {\bibfield  {journal} {\bibinfo  {journal} {Phys. Rev. Appl.}\
  }\textbf {\bibinfo {volume} {11}},\ \bibinfo {pages} {044083} (\bibinfo
  {year} {2019})}\BibitemShut {NoStop}%
\bibitem [{\citenamefont {Chen}\ and\ \citenamefont {Lidar}(2020)}]{Chen20}%
  \BibitemOpen
  \bibfield  {author} {\bibinfo {author} {\bibfnamefont {H.}~\bibnamefont
  {Chen}}\ and\ \bibinfo {author} {\bibfnamefont {D.~A.}\ \bibnamefont
  {Lidar}},\ }\href@noop {} {\bibfield  {journal} {\bibinfo  {journal} {Phys.
  Rev. Appl.}\ }\textbf {\bibinfo {volume} {14}},\ \bibinfo {pages} {014100}
  (\bibinfo {year} {2020})}\BibitemShut {NoStop}%
\bibitem [{\citenamefont {Seki}\ and\ \citenamefont
  {Nishimori}(2012)}]{Seki12}%
  \BibitemOpen
  \bibfield  {author} {\bibinfo {author} {\bibfnamefont {Y.}~\bibnamefont
  {Seki}}\ and\ \bibinfo {author} {\bibfnamefont {H.}~\bibnamefont
  {Nishimori}},\ }\href@noop {} {\bibfield  {journal} {\bibinfo  {journal}
  {Phys. Rev. E}\ }\textbf {\bibinfo {volume} {85}},\ \bibinfo {pages} {051112}
  (\bibinfo {year} {2012})}\BibitemShut {NoStop}%
\bibitem [{\citenamefont {Hormozi}\ \emph {et~al.}(2017)\citenamefont
  {Hormozi}, \citenamefont {Brown}, \citenamefont {Carleo},\ and\ \citenamefont
  {Troyer}}]{Hormozi17}%
  \BibitemOpen
  \bibfield  {author} {\bibinfo {author} {\bibfnamefont {L.}~\bibnamefont
  {Hormozi}}, \bibinfo {author} {\bibfnamefont {E.~W.}\ \bibnamefont {Brown}},
  \bibinfo {author} {\bibfnamefont {G.}~\bibnamefont {Carleo}},\ and\ \bibinfo
  {author} {\bibfnamefont {M.}~\bibnamefont {Troyer}},\ }\href@noop {}
  {\bibfield  {journal} {\bibinfo  {journal} {Phys. Rev. B}\ }\textbf {\bibinfo
  {volume} {95}},\ \bibinfo {pages} {184416} (\bibinfo {year}
  {2017})}\BibitemShut {NoStop}%
\bibitem [{\citenamefont {Nishimori}\ and\ \citenamefont
  {Takada}(2017)}]{Nishimori17}%
  \BibitemOpen
  \bibfield  {author} {\bibinfo {author} {\bibfnamefont {H.}~\bibnamefont
  {Nishimori}}\ and\ \bibinfo {author} {\bibfnamefont {K.}~\bibnamefont
  {Takada}},\ }\href@noop {} {\bibfield  {journal} {\bibinfo  {journal}
  {Frontiers in ICT}\ }\textbf {\bibinfo {volume} {4}},\ \bibinfo {pages} {2}
  (\bibinfo {year} {2017})}\BibitemShut {NoStop}%
\bibitem [{\citenamefont {Ohzeki}(2017)}]{Ohzeki17}%
  \BibitemOpen
  \bibfield  {author} {\bibinfo {author} {\bibfnamefont {M.}~\bibnamefont
  {Ohzeki}},\ }\href@noop {} {\bibfield  {journal} {\bibinfo  {journal} {Sci.
  Rep.}\ }\textbf {\bibinfo {volume} {7}},\ \bibinfo {pages} {41186} (\bibinfo
  {year} {2017})}\BibitemShut {NoStop}%
\bibitem [{\citenamefont {Crosson}\ \emph {et~al.}(2014)\citenamefont
  {Crosson}, \citenamefont {Farhi}, \citenamefont {Lin}, \citenamefont {Lin},\
  and\ \citenamefont {Shor}}]{Crosson14}%
  \BibitemOpen
  \bibfield  {author} {\bibinfo {author} {\bibfnamefont {E.}~\bibnamefont
  {Crosson}}, \bibinfo {author} {\bibfnamefont {E.}~\bibnamefont {Farhi}},
  \bibinfo {author} {\bibfnamefont {C.}~\bibnamefont {Lin}}, \bibinfo {author}
  {\bibfnamefont {H.-H.}\ \bibnamefont {Lin}},\ and\ \bibinfo {author}
  {\bibfnamefont {P.}~\bibnamefont {Shor}},\ }\href@noop {} {}\bibinfo
  {howpublished} {e-print arXiv:1401.7320} (\bibinfo {year} {2014})\BibitemShut
  {NoStop}%
\bibitem [{\citenamefont {Mandrà}\ \emph {et~al.}(2017)\citenamefont
  {Mandrà}, \citenamefont {Zhu},\ and\ \citenamefont {Katzgraber}}]{Mandra17}%
  \BibitemOpen
  \bibfield  {author} {\bibinfo {author} {\bibfnamefont {S.}~\bibnamefont
  {Mandrà}}, \bibinfo {author} {\bibfnamefont {Z.}~\bibnamefont {Zhu}},\ and\
  \bibinfo {author} {\bibfnamefont {H.~G.}\ \bibnamefont {Katzgraber}},\
  }\href@noop {} {\bibfield  {journal} {\bibinfo  {journal} {Phys. Rev. Lett.}\
  }\textbf {\bibinfo {volume} {118}},\ \bibinfo {pages} {070502} (\bibinfo
  {year} {2017})}\BibitemShut {NoStop}%
\bibitem [{\citenamefont {Barahona}(1982)}]{Barahona82}%
  \BibitemOpen
  \bibfield  {author} {\bibinfo {author} {\bibfnamefont {F.}~\bibnamefont
  {Barahona}},\ }\href@noop {} {\bibfield  {journal} {\bibinfo  {journal} {J.
  Phys. A: Math. Gen.}\ }\textbf {\bibinfo {volume} {15}},\ \bibinfo {pages}
  {3241} (\bibinfo {year} {1982})}\BibitemShut {NoStop}%
\bibitem [{\citenamefont {Choi}(2008)}]{Choi08}%
  \BibitemOpen
  \bibfield  {author} {\bibinfo {author} {\bibfnamefont {V.}~\bibnamefont
  {Choi}},\ }\href@noop {} {\bibfield  {journal} {\bibinfo  {journal} {Quantum
  Inf. Process.}\ }\textbf {\bibinfo {volume} {7}},\ \bibinfo {pages} {193}
  (\bibinfo {year} {2008})}\BibitemShut {NoStop}%
\bibitem [{\citenamefont {Choi}(2020)}]{Choi20}%
  \BibitemOpen
  \bibfield  {author} {\bibinfo {author} {\bibfnamefont {V.}~\bibnamefont
  {Choi}},\ }\href@noop {} {\bibfield  {journal} {\bibinfo  {journal} {Quantum
  Inf. Process.}\ }\textbf {\bibinfo {volume} {19}},\ \bibinfo {pages} {90}
  (\bibinfo {year} {2020})}\BibitemShut {NoStop}%
\bibitem [{\citenamefont {Amin}(2009)}]{Amin09}%
  \BibitemOpen
  \bibfield  {author} {\bibinfo {author} {\bibfnamefont {M.~H.~S.}\
  \bibnamefont {Amin}},\ }\href@noop {} {\bibfield  {journal} {\bibinfo
  {journal} {Phys. Rev. Lett.}\ }\textbf {\bibinfo {volume} {1102}},\ \bibinfo
  {pages} {220401} (\bibinfo {year} {2009})}\BibitemShut {NoStop}%
\bibitem [{\citenamefont {Amin}\ and\ \citenamefont {Choi}(2009)}]{Amin09b}%
  \BibitemOpen
  \bibfield  {author} {\bibinfo {author} {\bibfnamefont {M.~H.~S.}\
  \bibnamefont {Amin}}\ and\ \bibinfo {author} {\bibfnamefont {V.}~\bibnamefont
  {Choi}},\ }\href@noop {} {\bibfield  {journal} {\bibinfo  {journal} {Phys.
  Rev. A}\ }\textbf {\bibinfo {volume} {80}},\ \bibinfo {pages} {062326}
  (\bibinfo {year} {2009})}\BibitemShut {NoStop}%
\bibitem [{\citenamefont {Choi}(2022)}]{Choi22}%
  \BibitemOpen
  \bibfield  {author} {\bibinfo {author} {\bibfnamefont {V.}~\bibnamefont
  {Choi}},\ }\href@noop {} {}\bibinfo {howpublished} {e-print
  arXiv:2105.02110v2} (\bibinfo {year} {2022})\BibitemShut {NoStop}%
\bibitem [{\citenamefont {Braida}\ and\ \citenamefont
  {Martiel}(2021)}]{Braida21}%
  \BibitemOpen
  \bibfield  {author} {\bibinfo {author} {\bibfnamefont {A.}~\bibnamefont
  {Braida}}\ and\ \bibinfo {author} {\bibfnamefont {S.}~\bibnamefont
  {Martiel}},\ }\href@noop {} {\bibfield  {journal} {\bibinfo  {journal}
  {Quantum Inf. Process.}\ }\textbf {\bibinfo {volume} {20}},\ \bibinfo {pages}
  {260} (\bibinfo {year} {2021})}\BibitemShut {NoStop}%
\bibitem [{\citenamefont {Fujii}\ \emph {et~al.}(2022)\citenamefont {Fujii},
  \citenamefont {Komuro}, \citenamefont {Okudaira}, \citenamefont {Narita},\
  and\ \citenamefont {Sawada}}]{Fujii22}%
  \BibitemOpen
  \bibfield  {author} {\bibinfo {author} {\bibfnamefont {T.}~\bibnamefont
  {Fujii}}, \bibinfo {author} {\bibfnamefont {K.}~\bibnamefont {Komuro}},
  \bibinfo {author} {\bibfnamefont {Y.}~\bibnamefont {Okudaira}}, \bibinfo
  {author} {\bibfnamefont {R.}~\bibnamefont {Narita}},\ and\ \bibinfo {author}
  {\bibfnamefont {M.}~\bibnamefont {Sawada}},\ }\href@noop {} {}\bibinfo
  {howpublished} {e-print arXiv:2202.05927v1} (\bibinfo {year}
  {2022})\BibitemShut {NoStop}%
\bibitem [{\citenamefont {Dickson}(2011)}]{Dickson11}%
  \BibitemOpen
  \bibfield  {author} {\bibinfo {author} {\bibfnamefont {N.~G.}\ \bibnamefont
  {Dickson}},\ }\href@noop {} {\bibfield  {journal} {\bibinfo  {journal} {New
  J. Phys.}\ }\textbf {\bibinfo {volume} {13}},\ \bibinfo {pages} {073011}
  (\bibinfo {year} {2011})}\BibitemShut {NoStop}%
\bibitem [{\citenamefont {Wilkinson}(1989)}]{Wilkinson89}%
  \BibitemOpen
  \bibfield  {author} {\bibinfo {author} {\bibfnamefont {M.}~\bibnamefont
  {Wilkinson}},\ }\href@noop {} {\bibfield  {journal} {\bibinfo  {journal} {J.
  Phys. A: Mathematical and General}\ }\textbf {\bibinfo {volume} {22}},\
  \bibinfo {pages} {2795} (\bibinfo {year} {1989})}\BibitemShut {NoStop}%
\end{thebibliography}%

\end{document}